# Supplemental Transmission Aided Attenuation Correction for Quantitative Cardiac PET/MR

Mi-Ae Park, Vlad G. Zaha, Ramsey D. Badawi, and Spencer L. Bowen

*Abstract*— **Quantitative PET attenuation correction (AC) for combined cardiac PET/MR is a challenging problem. We propose and evaluate an AC approach that uses coincidences from a relatively weak and physically fixed sparse external source, in combination with that from the patient, to correct for PET attenuation based on physics principles alone. The low 30 ml volume of the source makes it easy to fill and place, and the method does not use prior image data or attenuation map assumptions. Our supplemental transmission aided maximum likelihood reconstruction of attenuation and activity (sTX-MLAA) algorithm contains an attenuation map update that maximizes the likelihood of terms representing coincidences originating from tracer in the patient and a weighted expression of counts segmented from the external source alone. Both external source and patient scatter and randoms are fully corrected. We evaluated performance of sTX-MLAA compared to reference standard CT-based AC with FDG PET/CT phantom studies; including modeling a patient with myocardial inflammation. Through an ROI analysis we measured <5% bias in activity concentrations for PET images generated with sTX-MLAA relative to CT-AC. PET background variability (from noise and sparse sampling) was substantially reduced with sTX-MLAA compared to using coincidences segmented from the transmission source alone for AC. The study suggests that sTX-MLAA will produce PET images on PET/MR with quantification comparable to PET/CT results during human cardiac exams.**

## I. INTRODUCTION

Combined PET/MR enables a comprehensive functional (mainly with PET) and anatomical (through MRI) workup of patients with cardiac diseases. Quantitative PET metrics provide benefits over visual analysis alone. For instance, PET/MR could aid in the evaluation of nonischemic myocardial inflammation [1]–[3]. Compared to patient workup with separate PET/CT and MR exams, cardiac PET/MR enables intrinsic temporal and spatial registration of PET and MR images (critical for diseases, such as cardiac sarcoidosis, that present with focal PET and MR contrast [4]) and reduces patient burden by removing one of two imaging studies [5].

Correcting for patient-related factors is necessary to minimize artifacts and produce quantitative PET images during cardiac PET/MR. Failure to correct for 511 keV photon attenuation typically has the largest impact, of all data corrections, on PET quantification. However, current MR-based attenuation correction (AC) methods have significant performance and/or practical limitations in cardiothoracic imaging. For instance, 50% of patients had μ-maps with MR susceptibility artifacts in one study [6], frequently due to cardiac stents (resulting in differences of up to 200% in SUVs relative to AC with manually corrected μ-maps). Another challenge for MR-AC is due to respiratory mismatch between the MR (typically breath hold) and PET (shallow breathing) exams, which can result in PET artifacts throughout the thoracic cavity; observed in 18 of 20 patients imaged on PET/MR [6].

As an alternative to MR-AC, methods have been developed that utilize the PET signal to reconstruct patient μ-maps. Algorithms using counts from the patient alone are termed maximum likelihood reconstruction of activity and attenuation (MLAA). However, MLAA methods suffer cross-talk between emission image and μ-map estimates, only reconstruct attenuation values up to a constant, and frequently rely on CT images for scatter correction (not practical on PET/MR); leading to PET image bias over 10% relative to CT-AC [7]. To address these limitations, researchers have used the intrinsic $^{176}$Lu radiation from PET detectors (i.e. alone or to produce an initial μ-map for MLAA) [8] or events from a hollow cylinder-shaped source alone [9] to reconstruct μ-maps. However, the relatively low count rate from $^{176}$Lu, or the cylindrical source, resulted in PET bias exceeding 5% in thoracic ROIs. In addition, filling and placing a hollow cylinder-shaped source that occupies the full PET field of view (FOV) is challenging.

Panin et al [10] developed an algorithm that uses counts from a rotating rod source, combined with those from the tracer in the patient, to correct for the deficits of MLAA. The rationale is that the higher count rate, for coincidences from the patient, combined with the more accurate attenuation estimation using coincidences from the external source, can produce AC PET images with higher signal-to-noise ratio (SNR) and accuracy

Funding was provided in part from the National Institutes of Health (grant number NIH-NIBIB R03EB028946-01A1). *Asterisk indicates corresponding author.*
M. Park is with the Department of Radiology, UT Southwestern Medical Center, Dallas, TX 75244 USA (email: Mi-Ae.Park@UTSouthwestern.edu).
V.G. Zaha is with the Cardiology Division, Department of Internal Medicine, the Advanced Imaging Research Center, and the Harold C. Simmons Cancer Center, UT Southwestern Medical Center, Dallas, TX 75244 USA (email: Vlad.Zaha@UTSouthwestern.edu).
R.D. Badawi is with the Department of Radiology, University of California Davis Medical Center, Sacramento, CA 95817 USA (email: rdbadawi@ucdavis.edu).
*S.L. Bowen is with the Department of Radiology, UT Southwestern Medical Center, Dallas, TX 75244 USA (email: Spencer.Bowen@UTSouthwestern.edu)

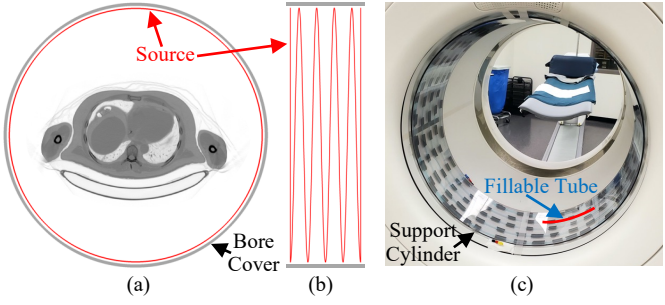

Fig. 1. Overview of the sparse transmission source. Schematic of the tori-helix geometry in the (a) transverse and (b) sagittal planes, fused with a CT image from a human patient for reference. (c) The source prototype placed in the bore of the PET/CT used for this study.

compared to using either strategy alone. The method of Panin et al, however, utilized PET data from 10 hr exams to correct for scatter from the external source (not practical for patient PET/MR exams). Furthermore, tight space confines of whole-body PET/MR scanners prevent integration of a rotating transmission source. We propose a supplemental transmission (TX) aided MLAA (sTX-MLAA) approach that uses 1) a physically fixed source, placed near the bore cover, that can be relatively easily filled and placed, 2) a transmission update that combines counts from the patient and external source, and 3) data corrections for TX source and patient scatter, to estimate and apply AC using PET data acquired during the exam intended for correction alone. The purpose of this study is to develop and evaluate sTX-MLAA with anthropomorphic phantom experiments approximating cardiac PET exams.

## II. METHODS

### A. Overview

Phantom acquisitions were performed on a time-of-flight (TOF) capable whole-body PET/CT; Siemens Biograph mCT Flow (Siemens Medical Solutions USA, Inc). We used a PET/CT to facilitate a direct comparison with the reference standard: PET images reconstructed with CT-AC. Our approach uses a custom TX source, detailed in section II.B. We then describe the supplemental transmission aided attenuation map reconstruction algorithm, including its theory (section II.C.1), PET data pre-processing (section II.C.2), method for updating and estimating the emission images (section II.C.3), approach for μ-map updates (section II.C.4), the combined update equation (section II.C.5), and data corrections (section II.C.6). Phantom experiments, data processing, and analysis are detailed in section II.D. $^{18}$F-fluorodeoxyglucose (FDG) was used for filling the TX source and phantoms in all experiments.

### B. Sparse transmission source

A sparse transmission source was constructed by wrapping a polytetrafluoroethylene (PTFE) fillable tube (ID=1.6 mm) around the outside of a hollow cylinder-shaped support (OD=76.2 cm, ID=75.2 cm, height=25.4 cm), made of polycarbonate. Polycarbonate annuli (ID=76.2 cm, OD=78 cm) were fitted on both ends of the support cylinder to center the TX source in the transverse FOV. Fig. 1 shows the prototype setup. We note that this configuration only reduces the ID of the Biograph mCT bore cover by 2.8 cm, which creates minimal mechanical interference when imaging patients that occupy a large fraction of the FOV. Additionally, by placing the source near the bore cover the interference between coincidences originating from the TX source and patient is minimized.

The geometry of the fillable tube was chosen as a tradeoff between tomographic sampling, the TX source count rate, impact on tracer reconstruction SNR, and ease of filling. Specifically, simulations demonstrated that sparse geometries can produce a higher noise equivalent count rate (surrogate for image SNR[2]) compared to rotating rods or a uniformly filled hollow cylinder-shaped source, at equal TX singles flux [11], [12]. Thus, we utilized two tori, near the axial ends of the PET FOV, joined with a four turn helix (pitch=5.3 cm), over a total axial length of 22 cm (PET axial FOV=22.1 cm). End-plane tori assist in reducing limited sampling artifacts that occur when using a helix alone. The PTFE tube had a total length of ~14.4 m, fill volume of 30 ml, and was fitted with Luer adapters (syringe connections) on the ends for easy filling and draining.

### C. Supplemental transmission aided attenuation map reconstruction

#### 1) Theory

The expected prompts $\bar{y}_{it}(\mu, x)$, as a function of the attenuation map ($\mu$) and patient radiotracer ($x$) image estimates, when scanning with a TX source, is given by

$$\bar{y}_{it}(\mu, x) = \left[b_{it} + \frac{\sum_j P_{itj} x_j}{N_i}\right] a_i + \frac{(\bar{s}_{it}(b) + \bar{s}_{it}(x))}{N_i} + \bar{r}_i \quad (1)$$

$$a_i = \exp(-l_i) = \exp\left(-\sum_j A_{ij}\mu_j\right) \quad (2)$$

where $b_{it}$ is the measured blank projection counts from the TX source (acquired without the patient in the PET scanner) for sinogram 3D coordinate $i$ and TOF index $t$, $j$ is the image voxel index, $P_{itj}$ are the elements of the TOF projection matrix, $N_i$ are the normalization sinogram factors, $a_i$ is the set of attenuation factors as a function of 3D sinogram coordinate alone, $\bar{s}_{it}(b)$ is a sinogram of expected scattered counts from the TX source only, $\bar{s}_{it}(x)$ represents expected scatter attributed to tracer in the patient, $\bar{r}_i$ is a sinogram of the random coincidence estimates (constant for all TOF indices at a given line-of-response $i$), $A_{ij} = \sum_t P_{itj}$ are elements of the non-TOF projection matrix, and $l_i$ are the product of projections lengths and linear attenuation coefficients (LACs). We note that $x$ accounts only for radiotracer in the patient alone (i.e. excluding the external source). For convenience, we refer to $x$ as the emission (EM) image. It is assumed that the blank projection ($b_{it}$) has been corrected for radionuclide decay, dead-time, and scan duration to match the scan with the patient.

Separately, by applying sinogram radial thresholding (implemented in section II.C.2), we can estimate prompts originating from the TX source alone ($\bar{y}_i^{\text{TX}_{\text{sep}}}$), as originally proposed by Mollet et al [9]. Specifically, using the mask, $m_{it}^{\text{TX}_{\text{sep}}}$, we sum TOF bins at sinogram projection elements that

spatially intersect the external TX source. Expected counts are described as follows:

$$\bar{y}_i^{\text{TXsep}}(\mu) = b_i^{\text{TXsep}} a_i + \frac{\bar{s}_i(b^{\text{TXsep}})}{N_i} + \bar{r}_i^{\text{TXsep}} \quad (3)$$

where $b_i^{\text{TXsep}}$ is the blank sinogram processed with the radial thresholding operation ($b_i^{\text{TXsep}} = \sum_t m_{it}^{\text{TXsep}} b_{it}$), $\bar{s}_i(b^{\text{TXsep}})$ are elements of the estimated scatter for segmented TX counts alone, and $\bar{r}_i^{\text{TXsep}}$ are expected randoms for the radially thresholded coincidences. The overall assumption is that net true coincidences (sum of trues and scatters) in $\bar{y}_i^{\text{TXsep}}$ are only counts from the TX source impacted by the µ-map, and there is negligible interference from patient annihilation photons.

To estimate the µ-map and EM image ($x$) we use the log-likelihood of the PET data. The equation combines 1) measured prompts from the patient and the TX source ($y_{it}$) and 2) the radially thresholded counts from the TX source alone ($y_i^{\text{TXsep}}$). Assuming the PET data represent independent Poisson random variables, the complete log-likelihood equation is given by

$$L(y|\mu, x) = \left[\sum_i \sum_t y_{it} \log(\bar{y}_{it}(\mu, x)) - \bar{y}_{it}(\mu, x)\right]$$
$$+ \alpha \left[\sum_i y_i^{\text{TXsep}} \log\left(\bar{y}_i^{\text{TXsep}}(\mu)\right) \quad (4)\right.$$
$$\left. - \bar{y}_i^{\text{TXsep}}(\mu)\right]$$

where $\alpha$ is a hyperparameter that adjusts the strength of the term in (4) that represents the likelihood of radially thresholded counts from the TX source. The $\alpha$ hyperparameter balances the impact of counts originating from the patient and those from the TX source; compensating for the sparse tomographic sampling and typically lower count rates, relative to the patient, of the TX source. Panin et al [10] optimized (4) with $\alpha = 0$, but as they found and as we show in this report, this is often inadequate for accurate AC with a physically fixed and sparse TX source.

To estimate the EM image ($x$) and µ-map ($\mu$) that maximizes the likelihood in (4), we developed an iterative algorithm that alternatively updates tracer and µ-map values. We detail these update methods in the following sections.

### 2) Data pre-processing

To estimate coincidences originating from the TX source and patient alone we employed a radial thresholding mask ($m_{it}^{\text{Radial}}$) in sinogram space, described as follows:

$$m_{it}^{\text{Radial}}(d^{\text{L}}, d^{\text{U}}) = \begin{cases} 1, & d^{\text{L}} \leq d(i,t) \leq d^{\text{U}} \\ 0, & \text{otherwise} \end{cases} \quad (5)$$

$$d(i,t) = \sqrt{d^{\text{D}} \sin\left(\left[i\%n_p - p_0\right]\Delta p / d^{\text{D}}\right)^2 + \left[(t - t_0)\Delta t\right]^2} \quad (6)$$

where $d^{\text{L}}$ and $d^{\text{U}}$ are the upper and lower radial limits, respectively, $d^{\text{D}}$ is the average distance from the center FOV to the detector, % is the modulo operator, $n_p$ is the number of radial projection bins, $\Delta p$ is the projection bin size, $p_0$ and $t_0$ are central bins, and $\Delta t$ the TOF sampling in space. Masks to segment coincidences originating from the TX source ($m_{it}^{\text{TXsep}}$) or patient alone ($m_{it}^{\text{EMsep}}$) were produced as follows:

$$m_{it}^{\text{TXsep}} = m_i^{\text{Triple}} m_{it}^{\text{Radial}}(340 \text{ mm}, 480 \text{ mm}) \quad (7)$$
$$m_{it}^{\text{EMsep}} = m_{it}^{\text{Radial}}(0 \text{ mm}, 340 \text{ mm}) \quad (8)$$

where $m_i^{\text{Triple}}$ is a mask used to employ the "triple-point" method; reducing the impact of scattered and randoms coincidences from the TX source. The mask selects lines-of-response (LORs) that intersect the TX source, and was produced by intensity thresholding the blank sinogram. Radial limits were chosen largely to minimize interference of counts from the patient on those from the transmission source, as these were typically much higher in magnitude. As the TX source was placed at a radius of 38.3 cm from the center FOV, these limits prioritize TX coincidences further from the patient. Segmented coincidences were produced with the masks as follows:

$$y_{it}^{\text{EMsep}} = m_{it}^{\text{EMsep}} y_{it}, \quad b_{it}^{\text{EMsep}} = m_{it}^{\text{EMsep}} b_{it} \quad (9)$$

$$y_i^{\text{TXsep}} = \sum_t m_{it}^{\text{TXsep}} y_{it}, \quad b_i^{\text{TXsep}} = \sum_t m_{it}^{\text{TXsep}} b_{it}, \quad (10)$$
$$\bar{r}_i^{\text{TXsep}} = \bar{r}_i \sum_t m_{it}^{\text{TXsep}}$$

where $y_{it}^{\text{EMsep}}$ and $b_{it}^{\text{EMsep}}$ are measured prompt and blank counts, respectively, segmented around the patient, and TX$_{\text{sep}}$ superscript sinograms are segmented around the TX source.

### 3) Emission reconstruction

For EM image reconstructions and updates we used TOF-based Ordinary Poisson ordered subsets expectation maximization (OP-OSEM), given by

$$\hat{x}_j^{n,1} = \hat{x}_j^{n-1}, \quad \hat{x}_j^0 = \hat{x}_j^{\text{Input}}$$

**for** m = 1, ···, $n_s$

$$\bar{y}_{it}^{\text{EMsep}} = a_i \left(\sum_j P_{itj} \hat{x}_j^{n,m} + N_i b_{it}^{\text{EMsep}}\right) + \bar{s}_{it}(x^{\text{EMsep}})$$
$$+ N_i \bar{r}_i$$

$$\hat{x}_j^{n,m+1} = \frac{\hat{x}_j^{n,m}}{\sum_{it \in S_m} \frac{P_{itj}}{N_i}} \sum_{i \in S_m, t} P_{itj} \frac{y_{it}^{\text{EMsep}}}{\bar{y}_{it}^{\text{EMsep}}} \quad (11)$$

$$= \frac{\hat{x}_j^{n,m}}{\sum_{i \in S_m, t} \frac{P_{itj}}{N_i}} \sum_{i \in S_m, t} P_{itj} \Delta_{itj}^{n,m}$$

**end for**

$$\hat{x}_j^n = \hat{x}_j^{n, n_s + 1}$$

where $\hat{x}_j^{n,m}$ is the EM image estimate at iteration $n$ and subset $m$, $n_s$ is the number of subsets, $S_m$ includes all projection indices in subset $m$, and $\bar{s}_{it}(x^{\text{EMsep}})$ is scatter from the EM image alone. We observed a ring artifact at the edge of EM images when using (11). This was due to slow convergence of $\hat{x}_j^{n,m}$ voxels where counts from the TX source were much greater than that from the patient. To correct for this, we adjusted sinogram update factors ($\Delta_{itj}^{n,m}$) as follows:

$$\Delta_{itj}^{n,m} = \begin{cases} \Delta_{itj}^{n,m}, & \text{if } \frac{P_{itj} \hat{x}_j^{n,m}}{b_{it}^{\text{EMsep}}} > 0.1 \\ 0, & \text{otherwise} \end{cases} \quad (12)$$

where the equation was applied during each subset update.

*4) Transmission reconstruction*

To reconstruct μ-map estimates we employed a penalized-likelihood algorithm. The optimal attenuation image voxels ($\hat{\mu}$) or update factors were determined by maximizing the log-posterior density function below:

$$\hat{\mu}(y) = \underset{\mu \geq 0}{\mathrm{argmax}}[L^{\mathrm{sum}}(y|\mu,x) - \beta R(\mu)] \quad (13)$$

$$L^{\mathrm{sum}}(y|\mu,x) = \sum_i \big[y_i^{\mathrm{sum}} \log(\bar{y}_i^{\mathrm{sum}}(\mu,x)) \\
- \bar{y}_i^{\mathrm{sum}}(\mu,x)] \\
+ \alpha \big[y_i^{\mathrm{TXsep}} \log(\bar{y}_i^{\mathrm{TXsep}}(\mu)) \\
- \bar{y}_i^{\mathrm{TXsep}}(\mu)\big]\big] \quad (14)$$

where $y_i^{\mathrm{sum}} = \sum_t y_{it}$ and $\bar{y}_i^{\mathrm{sum}} = \sum_t \bar{y}_{it}$ are measured and expected prompts summed over all TOF bins, $\beta$ is a hyperparameter, and $R(\mu)$ is a roughness penalty function.

Optimization of (13) was performed with an ordered subset implementation of the separable paraboloidal surrogates (SPS) algorithm developed by Erdogan and Fessler [13]. A non-penalized SPS algorithm for combined emission and transmission MLAA, with $^{176}$Lu, was first proposed by Cheng et al [14]. The μ-map reconstruction algorithm is given by

$$\hat{\mu}_j^{n,1} = \hat{\mu}_j^{n-1}, \qquad \hat{u}_j^0 = \hat{\mu}_j^{\mathrm{Input}}$$

**for** m = 1, ⋯, $n_s$

$$\hat{\mu}_j^{n,m+1} = \hat{\mu}_j^{n,m} \\
+ \frac{n_S \sum_{i \in S_m} A_{ij} \left[\dot{h}_i^{\mathrm{sum}} + \alpha \dot{h}_i^{\mathrm{TXsep}}\right] - \beta \dot{R}(\hat{\mu}_j^{n,m})}{n_S \sum_{i \in S_m} A_{ij}\gamma_i \left[c_i^{\mathrm{sum}} + \alpha c_i^{\mathrm{TXsep}}\right] + \beta R_\psi(\hat{\mu}_j^{n,m})} \quad (15)$$

**end for**

$$\hat{u}_j^n = \hat{u}_j^{n,n_s+1}$$

where $\dot{R}(\hat{\mu}_j^k)$ is the derivative of the roughness penalty function, $R_\psi(\hat{\mu}_j^k)$ is a separable surrogate for the penalty function, $\gamma_i = \sum_j A_{ij}$, and $\dot{h}_i$ and $c_i$ are functions that both input a forward projection of the current μ-map estimate ($l_i^{n,m} = \sum_j A_{ij}\hat{\mu}_j^{n,m}$). These functions are defined further for $\dot{h}_i$

$$\dot{h}_i^{\mathrm{sum}} = \dot{h}(l_i^{n,m}, y_i^{\mathrm{sum}}, \bar{y}_i^{\mathrm{sum}}, \hat{b}_i^{\mathrm{sum}}), \\
\dot{h}_i^{\mathrm{TXsep}} = \dot{h}(l_i^{n,m}, y_i^{\mathrm{TXsep}}, \bar{y}_i^{\mathrm{TXsep}}, b_i^{\mathrm{TXsep}}) \quad (16)$$

$$\dot{h}(l_i^{n,m}, y_i, \bar{y}_i, \hat{b}_i) = \left(1 - \frac{y_i}{\bar{y}_i}\right)\hat{b}_i \exp(-l_i^{n,m}) \quad (17)$$

where $\hat{b}_i^{\mathrm{sum}} = \sum_t b_{it} + \frac{\sum_j A_{ij}\hat{x}_j^n}{N_i}$, using the current EM image reconstruction. The $c_i$ functions are defined as follows:

$$c_i^{\mathrm{sum}} = c(l_i^{n,m}, y_i^{\mathrm{sum}}, \bar{y}_i^{\mathrm{sum}}, \hat{b}_i^{\mathrm{sum}}, \bar{n}_i^{\mathrm{sum}}), \quad c_i^{\mathrm{TXsep}} = c(l_i^{n,m}, y_i^{\mathrm{TXsep}}, \bar{y}_i^{\mathrm{TXsep}}, b_i^{\mathrm{TXsep}}, \bar{n}_i^{\mathrm{TXsep}}) \quad (18)$$

$$c(l_i, y_i, \bar{y}_i, \hat{b}_i, \bar{n}_i) = \begin{cases} \frac{2}{(l_i)^2}\left\{\hat{b}_i[1-\exp(-l_i)] - y_i\log\left[\frac{\hat{b}_i+\bar{n}_i}{\bar{y}_i}\right] + l_i\left[\frac{y_i}{\bar{y}_i} - 1\right]\hat{b}_i\exp(-l_i)\right\}, & \text{if } l_i > 0 \\ \hat{b}_i\left(1 - \frac{\bar{n}_i y_i}{(\hat{b}_i+\bar{n}_i)^2}\right), & \text{otherwise} \end{cases} \quad (19)$$

where $\bar{n}_i$ is the sum of the expected randoms and scatter. These combined noise terms are given by

$$\bar{n}_i^{\mathrm{sum}} = \frac{(\bar{s}_i(b^{\mathrm{TXsum}}) + \sum_t \bar{s}_{it}(\hat{x}))}{N_i} + n_t \bar{r}_i \quad (20)$$

$$\bar{n}_i^{\mathrm{TXsep}} = \frac{\bar{s}_i(b^{\mathrm{TXsep}})}{N_i} + \bar{r}_i^{\mathrm{TXsep}} \quad (21)$$

where $\bar{s}_i(b^{\mathrm{TXsum}})$ is the scatter originating from the TX source summed over all TOF bins, $n_t$ is the number of TOF bins, and the terms in (21) are defined after (3). The update equations utilize optimal curvature, as represented by the $c_i$ term in (19) [13], to provide fast convergence for μ-map updates and the sTX-MLAA as a whole.

The roughness penalty function is defined as

$$R(\mu) = \sum_{j=1}^Q \sum_{k \in \mathcal{N}_j} \kappa_{jk} V(\mu_j - \mu_k) \quad (22)$$

where $\mathcal{N}_j$ is the 26-voxel 3D neighborhood around voxel $j$, $\kappa_{jk}$ is the inverse Euclidean distance between the voxel coordinates $j$ and $k$, and $V(\mu_j - \mu_k)$ is the edge preserving Huber function.

The Huber equation is given by

$$V(\mu_j - \mu_k) = \begin{cases} (\mu_j - \mu_k)^2/2/\delta, & \text{if } |\mu_j - \mu_k| \leq \delta \\ |\mu_j - \mu_k| - \delta/2, & \text{otherwise} \end{cases} \quad (23)$$

where $\delta$ was fixed at 3.0 x $10^{-2}$ cm$^{-1}$ for all experiments.

We also utilized segmented TX source data alone (TX$_{\mathrm{sep}}$) to reconstruct μ-maps, by modifying (15) as follows:

$$\hat{\mu}_j^{n,m+1} = \hat{\mu}_j^{n,m} + \frac{n_S \sum_{i \in S_m} A_{ij} \dot{h}_i^{\mathrm{TXsep}} - \beta \dot{R}(\hat{\mu}_j^{n,m})}{n_S \sum_{i \in S_m} A_{ij}\gamma_i c_i^{\mathrm{TXsep}} + \beta R_\psi(\hat{\mu}_j^{n,m})} \quad (24)$$

where the terms are defined above.

*5) sTX-MLAA combined update equation*

We combined sequential EM image and attenuation map updates, as shown in pseudo-code Algorithm 1, to reconstruct the final μ-map and nuisance EM image.

*6) Data corrections*

Scattered photons originating from the TX source and patient were corrected independently using two different algorithms. For both, the single scatter simulation (SSS) method [15] was employed to estimate unscaled scatter sinograms. Our key assumption for scatter correction was that there is negligible corruption of radially-thresholded scatter originating from the patient with that from the external source, and vice versa.

*a)  Transmission scatter*

For scatter estimation of events originating from the TX source we performed the following steps: 1) reconstruct the blank sinogram with a non-TOF implementation of OP-OSEM in (11), 2) reconstruct an initial µ-map using the radially-thresholded data ($y_i^{TX_{sep}}$) and the algorithm in (24), 3) estimate absolute (unscaled) scatter sinograms with a fully 3D implementation of the SSS method, using the reconstructed blank sinogram and initial µ-map as inputs, and 4) scale scatter estimates by tail fitting on the blank subtracted sinogram ($s_i^{TX_{sep}}$). This blank subtracted sinogram is expected to contain only scattered events for LORs intersecting the TX source but not the attenuating material, and is given by

$$s_i^{TX_{sep}} = N_i \left( y_i^{TX_{sep}} - b_i^{TX_{sep}} - \bar{r}_i^{TX_{sep}} \right). \quad (25)$$

To define the support of $s_i^{TX_{sep}}$ we multiplied elements of the "triple-point" mask ($m_i^{Triple}$), in (7), with those in a mask of "tail-regions" neighboring the attenuating material ($m_i^{Tail}$); defined using the current estimate of attenuation-correction factors ($\exp(\sum_j A_{ij} \hat{\mu}_j^n)$). To compute the scale factors, linear least squares fitting was performed between the raw SSS-estimated scatter distributions and the difference sinogram in the masked tail regions across all sinogram slices (621 total scale factors).

In sum, to estimate the scatter in radially-thresholded transmission sinograms, $\bar{s}_i(b^{TX_{sep}})$ in (3), we utilized the above scaling approach directly. To estimate the TX scatter summed over all the TOF bins, $\bar{s}_i(b^{TX_{sum}})$ in (20), we scaled $\bar{s}_i(b^{TX_{sep}})$ as follows:

$$\bar{s}_i(b^{TX_{sum}}) = \left( \frac{\sum_{it} m_i^{Triple} m_i^{Tail} b_{it}}{\sum_i m_i^{Tail} b_i^{TX_{sep}}} \right) \bar{s}_i(b^{TX_{sep}}) \quad (26)$$

which is equivalent to scaling by the sum of trues in the original blank sinogram, added across the TOF dimension, divided by that in the radially-thresholded blank sinogram.

*b)  Patient scatter, randoms, and dead-time*

To estimate scatter originating from the patient alone, $\bar{s}_{it}(x^{EM_{sep}})$ in (11) and $\bar{s}_{it}(\hat{x})$ in (20), we utilized absolute scatter estimation [16]. In this approach, the current EM image reconstruction ($\hat{x}_j^n$) and attenuation image were fed into a 2D implementation of the SSS. Reconstruction of the EM image and estimation of scatter with SSS was globally iterated a total of four times to refine the scatter estimate. The resulting 2D scatter sinogram was then converted to 3D with inverse single slice rebinning and used directly (i.e. without additional scaling) to produce $\bar{s}_{it}(\hat{x})$, or $\bar{s}_{it}(x^{EM_{sep}}) = m_{it}^{EM_{sep}} \bar{s}_{it}(\hat{x})$

Randoms were estimated with a variance reduction algorithm [17], while dead-time was corrected at the bucket level (combination of 4 contiguous detector blocks in a ring) using singles count rates, as in [18]. The full sTX-MLAA algorithm, including global iterations to correct for scatter, is shown in Algorithm 2.

---

Algorithm 1. sTX_MLAA($\hat{\mu}, \hat{x}, \alpha, \beta$, MLAAIter): Supplemental transmission-aided MLAA

1: **Input**: $\hat{\mu}, \hat{x}, \alpha, \beta$, and MLAAIter
2: OSEMIter=1, TXIter=1
3: **Initialize** $\hat{\mu}_j^0 = \hat{\mu}$ and $\hat{x}_j^0 = \hat{x}$
4: **for** $n = 1$ to MLAAIter **do**
5: $\quad a_i = \exp\left(-\sum_j A_{ij} \mu_j^{n-1}\right)$
6: $\quad$ **Compute** $\hat{x}_j^n$ with OSEM as in (11), by inputting $\hat{x}_j^{n-1}, a_i$, and OSEMIter iterations
7: $\quad \hat{b}_i^{sum} = b_i^{sum} + \frac{\sum_j A_{ij} \hat{x}_j^n}{N_i}$
8: $\quad$ **Compute** $\hat{\mu}_j^n$ with TX update in (15), by inputting $\mu_j^{n-1}, \alpha, \beta$, and TXIter iterations
9: **end for**
10: **return** final $\hat{\mu}$ and $\hat{x}$

---

### D. Phantom study

*1) Data acquisition and reconstruction*

We increased the coincidence window ($2\tau$) for all PET scans to maximize recovery of external source coincidences. Default $2\tau$ is 4.06 ns and vendor-produced sinograms have 13 TOF bins of 312 ps each. This corresponds to a maximum spatial TOF difference of 30.4 cm from the center FOV. Thus, default settings reduce the count rate of coincidences from the TX source (radial offset of 38.1 cm), as noted by Panin et al [10], largely impacting central sinogram bins. We used $2\tau$ =7.81 ns and performed offline binning of list-mode data to produce sinograms with up to 25 TOF bins of 312 ps each. After radial thresholding (section II.C.2), $2\tau$ was reduced to 6.25 ns (maximum radial offset of 48.0 cm) to produce measured prompts ($y_{it}$) in (4), and transmission (TX$_{sep}$) sinograms in (10).

All experiments were acquired at a single patient table position, mimicking the entirety or portion of a cardiac PET exam. For CT-AC, the imaging protocol included acquisition with 120 kVp and variable mA, and µ-maps were generated using the bilinear transform method implemented in the e7tools (Siemens Medical Solutions USA, Inc.). Transmission source blank acquisitions were always performed after phantom scanning, with an acquisition time of 45 min.

For sTX-MLAA we utilized the following fixed parameters: 1) 3 global scatter iterations (ScatterIter on line 2 of Algorithm 2), 2) 4 subsets for emission and transmission updates, 3) 20 iterations for the initial µ-map and tracer reconstructions (lines 4 and 9, respectively, in Algorithm 2), and 4) MLAAIter set to 60. Initial images were uniform cylinders, covering the full PET FOV (transverse D=69.7 cm, axial length=22.1 cm), set to the LAC of water (9.6 x 10$^{-2}$ cm$^{-1}$) or ones, for µ-map and EM images, respectively. Reconstructions had 400 x 400 x 109 voxels with sizes of 2.04 x 2.04 x 2.03 mm. Critically, for all experiments the scan duration of PET data input into sTX-MLAA, matched that input into the phantom PET image reconstructions where AC was applied (as described below).

Algorithm 2. Pseudo-code for the complete sTX-MLAA method with recursive scatter estimation

1: **Input**: $\alpha, \beta$, ScatterIter, and MLAAIter
2: **for** $p = 1$ to ScatterIter **do**
3:     **Initialize** $\hat{\mu}$ and $\hat{x}$ with uniform values
4:     **Initial** $\hat{\mu}$ reconstruction with $y_i^{TX_{sep}}$ in (24)
5:     **If** $p == 1$
6:         **Compute** patient scatter $(\bar{s}_{it}(\hat{x}), \bar{s}_{it}(x^{EM_{sep}}))$
7:         **Compute** TX scatter $(\bar{s}_{it}(b^{TX_{sep}}), \bar{s}_i(b^{TX_{sum}}))$
8:     **end if**
9:     **Initial** $\hat{x}$ reconstruction with $y_{it}^{EM_{sep}}$ in (11)
10:     $(\hat{\mu}, \hat{x}) := $ sTX_MLAA$(\hat{\mu}, \hat{x}, \alpha, \beta, $ MLAAIter$)$
11:     **Compute** patient $(\bar{s}_{it}(\hat{x}), \bar{s}_{it}(x^{EM_{sep}}))$ and TX $(\bar{s}_{it}(b^{TX_{sep}}), \bar{s}_i(b^{TX_{sum}}))$ scatter
12: **end for**
13: **return** final $\hat{\mu}$ and $\hat{x}$

To compare different AC methods, PET images were reconstructed with TOF OP-OSEM (no PSF modeling, 2 iterations, 21 subsets), followed by smoothing with a 5 mm FWHM Gaussian (200 x 200 x 109 matrix size with 4.07 x 4.07 x 2.03 mm voxels). Input sinograms were binned to 13 TOF bins, matching the vendor default. The reconstruction protocol followed that used for cardiac evaluations at our institution and the literature [19]. For consistency, only the μ-map was substituted with all other factors remaining constant (i.e. tracer reconstructions produced as part of sTX-MLAA were not evaluated). Besides AC, corrections for scatter (absolute scatter scaling), randoms, dead-time, and normalization were included. All reconstructions were implemented with the e7tools (Siemens Medical Solutions USA, Inc.), custom C-executables, and MATLAB (MathWorks).

*2) NEMA IEC body phantom experiments*

A NEMA IEC PET body phantom (Data Spectrum) was scanned in two configurations to approximate a human torso. The phantom contains a 1) fillable background and six spheres ranging in ID from 10 to 37 mm, and 2) a cylindrical "lung insert", filled only with polystyrene beads (i.e. no activity). FDG activity concentration ratio of all spheres to background was 4:1. The phantom was filled with a total activity and activity concentration comparable to the torso of a 70 kg adult (i.e. 370 MBq FDG injection, 60 min uptake, 20% excretion).

For the first study the phantom was imaged alone. At the start of imaging, total activity and the activity concentration in the fillable background were 28.3 MBq and 2.8 kBq/ml, respectively, and total activity in the TX source was 16.5 MBq. The phantom was scanned for a total of 30 min. PET list-mode was binned into 10 and 30 min acquisitions. These scan times are consistent with the cardiac focused table position, on PET/CT, and near the lower end for PET/MR. For sTX-MLAA, the penalty-strength, $\beta$ in (15), was set to $2^{12}$.

For the second setup, the phantom was supplemented with elements to reflect a patient with arms down and expand the range of LACs (Fig. 2). Two two-liter bottles were used for arms, while a Teflon rod insert (mimicking cortical bone) and an implantable cardioverter defibrillator (ICD) generator were positioned directly on the body phantom. At the start of imaging, total activity and the activity concentration in the torso fillable background were 20.6 MBq and 2.1 kBq/ml, respectively, and total activity in the TX source was 12.7 MBq. Acquisition time was 20 min. For sTX-MLAA, $\beta$ in (15) was set to $2^{14}$, and reconstructed μ-maps here were post-processed by setting voxels below $1.0 \times 10^{-2}$ cm$^{-1}$ to zero, to minimize impact from low-intensity noisy background regions.

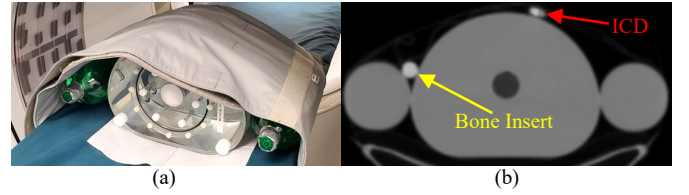

Fig. 2. Study setup for NEMA IEC body phantom with arms, a bone insert, and ICD generator. (a) Placement of the phantom on the PET/CT table. (b) CT μ-map with display limits of 0.0 to 0.2 cm$^{-1}$.

We quantified both absolute and relative quantification of the AC methods with an ROI analysis, conducted according to NEMA NU 2-2007 performance guidelines [20]. Briefly, ROIs with diameters matching the inner diameters of the fillable spheres were placed on a single transverse slice, and concentric ROIs of the same diameters were placed on the background at five slices. The position of ROIs was replicated for the different reconstructions of each configuration. Sphere and background absolute uptake (in kBq/ml), sphere contrast relative to the background, the cold lung insert relative to background, and background variability were calculated.

*3) Anthropomorphic torso phantom with cardiac insert*

An anthropomorphic torso phantom with a cardiac insert (Data Spectrum) was prepared to approximate a patient with active cardiac sarcoidosis injected with FDG. The phantom contains fillable compartments representing organs (e.g. lungs, liver, etc.), listed in Table I, as well as a segment of myocardial inflammation. Polystyrene beads and a Teflon plastic rod are used to approximate LACs of the lungs and spine, respectively.

We filled the phantom assuming the following: 1) patient was on a specialized diet (i.e. high-fat, high-protein, low-carbohydrate) to significantly reduce non-specific myocardial FDG uptake, 2) uptake time of 90 min and 20% tracer excretion, and 3) relative activity concentration ratios based on patient PET/CT data from our institution and prior reports [3], [21], as listed in Table I. At the start of imaging, total activity in the phantom was 26.1 MBq, background activity concentration was 1.9 kBq/ml, and TX source activity was 14.8 MBq. The phantom was scanned for 35 min, and images reconstructed for the full duration. For sTX-MLAA, $\beta$ in (15) was set to $2^{14}$.

TABLE I
FDG CONCENTRATION RATIOS FOR THE
ANTHROPOMORPHIC TORSO PHANTOM WITH CARDIAC INSERT

| Organ | Concentration Ratio (Normalized to Background) |
|---|---|
| Background | 1.0 |
| Liver | 2.7 |
| Lungs | 0.5 |
| Myocardium | 2.2 |
| Cardiac Defect | 4.9 |
| Ventricular Blood | 2.2 |

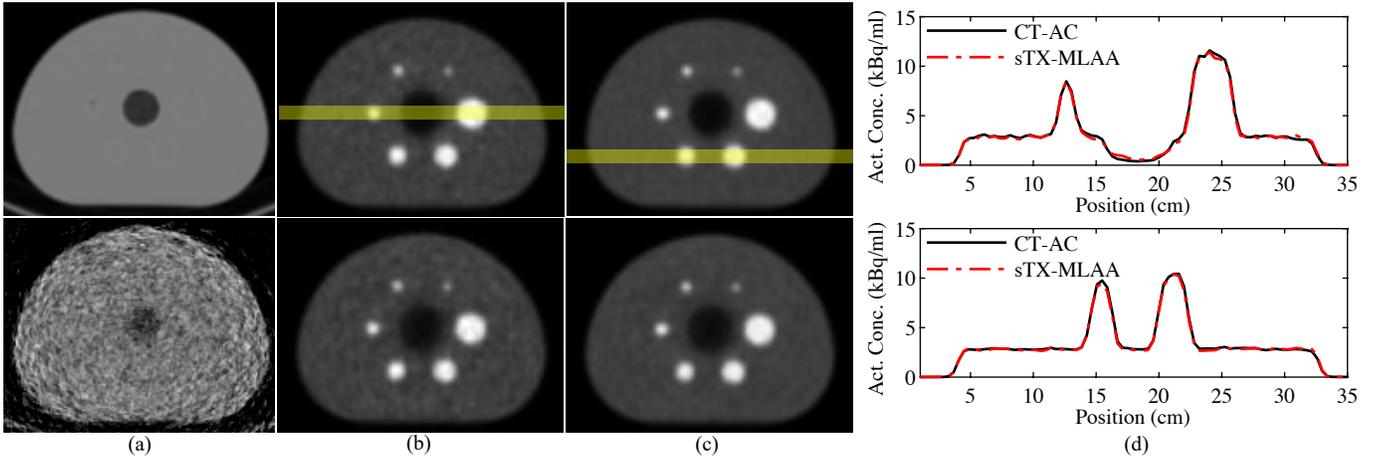

Fig. 3. IEC phantom μ-map and PET results for two acquisition times. Transverse images of (a) μ-maps from CT (top) and sTX-MLAA for a 10 min scan (display limits=0-0.2 cm$^{-1}$). (b,c) PET images generated with CT (top) and sTX-MLAA ($\alpha$=10) attenuation corrections for (b) 10 min and (c) 30 min scans. (d) Line profiles from the 10 min (top) and 30 min acquisitions, with locations shown in top panels (b) and (c), respectively.

## III. RESULTS

### A. NEMA IEC body phantom experiments

Example μ-map and PET images reconstructed with CT-based and sTX-MLAA attenuation corrections, for the IEC phantom alone, are shown in Fig. 3. Visually, PET images with AC using sTX-MLAA showed good agreement with CT-AC results. However, for sTX-MLAA PET images we did observe subtle increases in noise and/or high-frequency artifacts as well as a tendency to overestimate uptake for the cold lung insert.

Fig. 4 compares quantification between CT and sTX-MLAA attenuation corrected PET images at two acquisition times. As shown in Fig. 4a, the choice of $\alpha$ in (15), has a large impact on the root-mean-square error (RMSE) of sphere uptake bias and image noise. As $\alpha$ was increased from 1.0 to 17.5, sphere uptake bias decreased by ≥5% while background variability increased by ≥1%, for 10 and 30 min scans. We chose $\alpha$=10 as the optimum for this study, as bias decreased by <1% compared to the lowest value at the maximum $\alpha$ (17.5), while variability continued increasing past this point. Fig. 4b shows bias for sTX-MLAA PET images at $\alpha$=10. Sphere contrast, background contrast, and contrast recovery for sTX-MLAA were all within 4.9% of CT-AC results. Lung residual (ratio of lung to background uptake) increased from ~16.0%, with CT-AC, to 24.2% and 22.9%, with sTX-MLAA AC, for 10 and 30 min scans, respectively. Table II shows PET background variability at $\alpha$=10. The RMSE increase (relative to CT-AC) was 1.3% and 1.2% for 10 min and 30 min scans, respectively.

Fig. 5 compares performance of sTX-MLAA against reconstructing the μ-map with the SPS algorithm in (24) using coincidences segmented from the TX source alone. We increased the iteration number of the TX reconstruction (line 4 in Algorithm 2) to 80 in the final global scatter iteration ($p$ =ScatterIter) and then terminated algorithm execution (matching total TX iterations with sTX-MLAA). The penalty strength, $\beta$ in (24), was varied from $2^5$ to $2^{14}$ in reconstructions with segmented coincidences. sTX-MLAA produced PET images with substantially lower background variability (>3.6%) compared to transmission reconstructions with segmented external coincidences, at matched background bias.

Fig. 6 qualitatively compares AC methods for the IEC body phantom supplemented with arms, a bone insert, and ICD generator. The CT μ-map, at the same slice, is shown in Fig. 2b. We found $\alpha$=1, in (15), produced optimal results here. PET images reconstructed with CT-based and sTX-MLAA AC showed excellent visual agreement. However, we did again observe subtle increases in noise and/or high frequency artifacts in sTX-MLAA results, as well as positive bias in the lung insert. The phantom's left arm slightly extended past the transverse CT FOV (D=50 cm), possibly impacting CT-AC near this position.

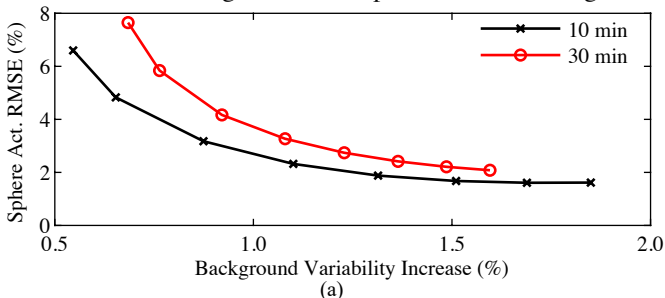

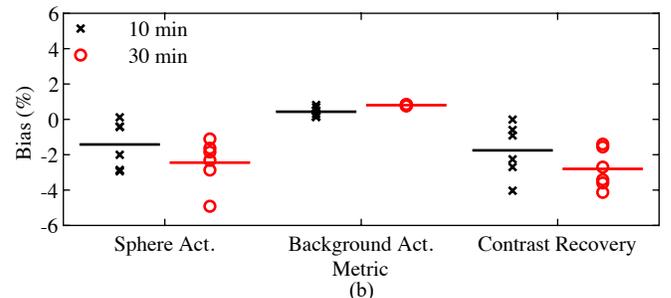

Fig. 4. Quantification of sTX-MLAA performance for the IEC phantom and two acquisition times. (a) RMSE in measured sphere activity concentration versus increase in background variability, with respect to CT-AC PET. Each data point is for a different $\alpha$ in (15); range=1.0-17.5. Both the RMSE in sphere uptake and increase in background variability are across all sphere ROI sizes. (b) Bias ($\alpha$=10), relative to CT-AC, in sphere activity concentration, background activity concentration, and contrast recovery. Each data point is a unique sphere ROI size and lines denote mean values.

TABLE II
BACKGROUND VARIABILITY IN NEMA IEC BODY PHANTOM
PET IMAGES WITH DIFFERENT AC METHODS

| Sphere ID | Background Variability (%) | | | |
|---|---|---|---|---|
| | 10 min | | 30 min | |
| | CT | sTX-MLAA | CT | sTX-MLAA |
| 10 mm | 4.1 | 6.2 | 3.0 | 4.8 |
| 13 mm | 3.9 | 5.4 | 2.7 | 3.9 |
| 17 mm | 3.4 | 4.4 | 2.3 | 3.4 |
| 22 mm | 2.9 | 3.8 | 2.0 | 2.9 |
| 28 mm | 2.5 | 3.4 | 1.8 | 2.8 |
| 37 mm | 2.0 | 2.9 | 1.6 | 2.6 |

Fig. 7 details the PET image ROI analysis for the IEC phantom with arms. Sphere and background contrast for sTX-MLAA were within 3.5% (0.1 kBq/ml) of CT-AC, while bias in sphere contrast recovery was ≤5.6% for all spheres. RMSE increase in background variability was 0.9%, and increase in the lung residual was 10.8%, for sTX-MLAA versus CT AC.

### B. Anthropomorphic torso phantom with cardiac insert

Fig. 8 shows a qualitative comparison between AC methods for the phantom study modeling a patient with active cardiac sarcoidosis. We used $\alpha$=10, in (15), for sTX-MLAA reconstructions. We noted high visual agreement between PET images reconstructed with CT and sTX-MLAA AC methods. As confirmed by line profile analysis (Fig. 8c), uptake in all body regions (notably the cardiac defect and lung compartments) matched closely between the two AC methods, although there was a slight overestimation in sTX-MLAA for the liver compared to CT-AC.

To quantify results further, ROIs were manually placed on transverse images of the different fillable compartments and mean uptake measured. A large spherical VOI (D=6 cm) was drawn on the liver. Table III details these quantitative findings. Uptake in PET images corrected with sTX-MLAA was within 3.7% (0.2 kBq/ml) of CT-AC results for all organs, with the difference in liver uptake the largest recorded. Notably, the activity concentration in the cardiac defect for sTX-MLAA was only 0.9% (0.05 kBq/ml) higher than CT-AC results.

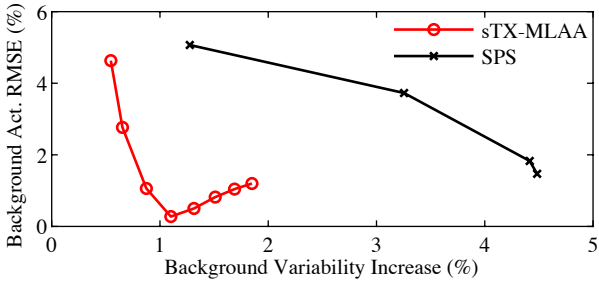

Fig. 5. Comparison of sTX-MLAA and transmission reconstruction inputting segmented TX data only (SPS), on PET attenuation correction for the IEC phantom; 10 min scan. RMSE in measured background activity concentration versus increase in background variability, with respect to CT-AC PET, for different $\alpha$ in sTX-MLAA and a range of $\beta$ in (24), for SPS.

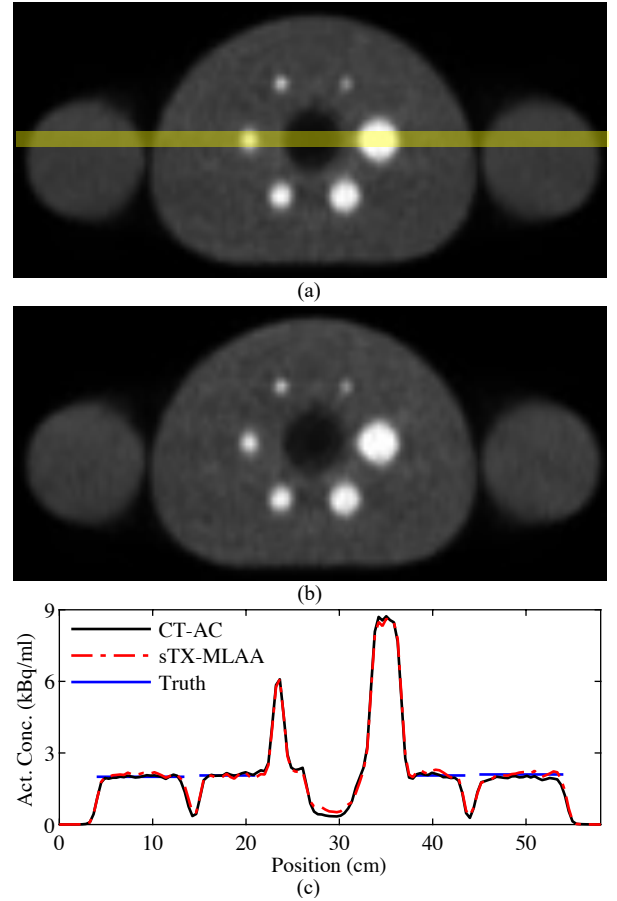

Fig. 6. Evaluation of sTX-MLAA performance for the IEC phantom with arms, bone insert, and ICD. Transverse PET images, reconstructed with (a) CT and (b) sTX-MLAA (using $\alpha$=1) attenuation correction. (c) Line profiles through the two PET reconstructions, with location shown in (a). The "Truth" profiles represent those measured with a dose calibrator. The phantom's left arm is at position near 50 cm.

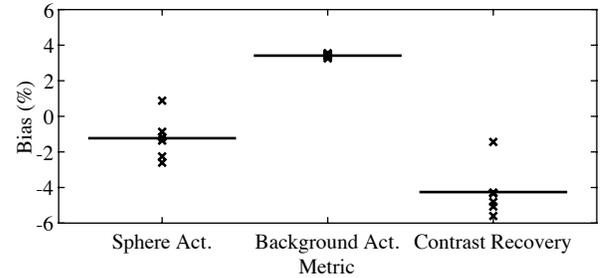

Fig. 7. sTX-MLAA quantification for the IEC phantom with arms, bone insert, and ICD, using α=1. Bias, with respect to CT-AC results, in sphere activity concentration, background activity concentration, and contrast recovery. Each data point represents a different sphere ROI size and lines denote mean values.

## IV. DISCUSSION

ROI results demonstrated that uptake measured in PET images corrected for attenuation with sTX-MLAA were within 5% of CT-AC results, for all anthropomorphic phantom studies. sTX-MLAA PET images had lower uptake bias than PET images generated with AC using coincidences largely originating from the patient alone, as indicated by quantitative results at comparatively low $\alpha$ values (Fig. 4a); approximating

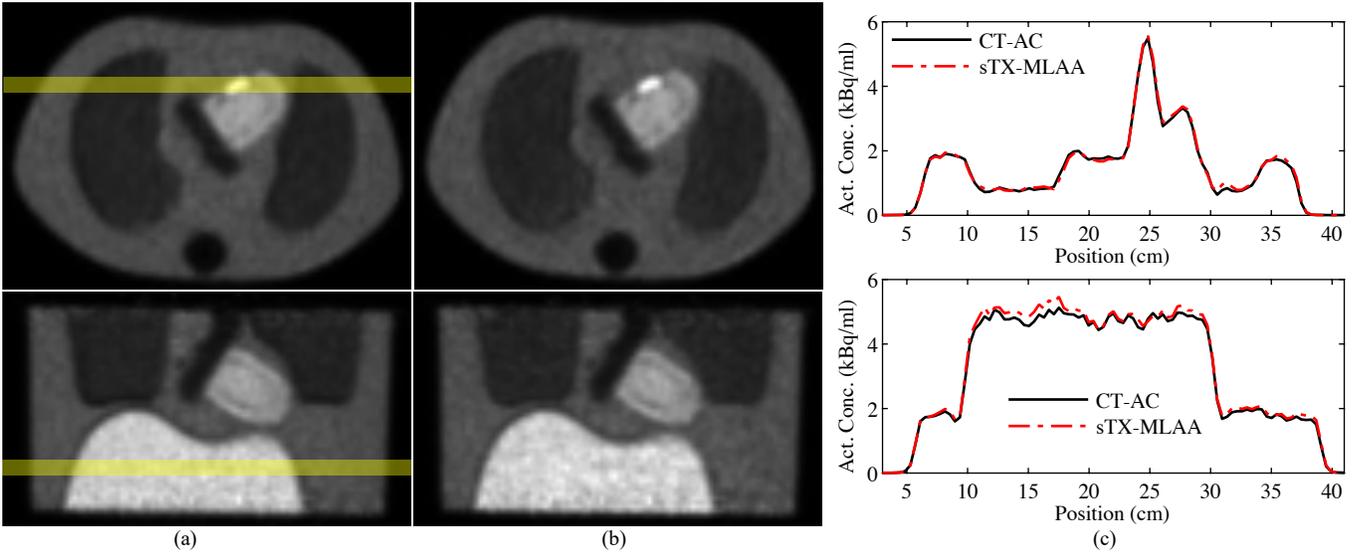

Fig. 8. Evaluation of sTX-MLAA performance for an anthropomorphic phantom with a cardiac insert. Transverse (top) and coronal PET images, reconstructed with (a) CT and (b) sTX-MLAA (using $\alpha$=10) attenuation correction. (c) Line profiles through PET images with the two AC methods for transverse (top) and coronal slices, with locations shown in top and bottom panels of (a), respectively.

MLAA without a μ-map intensity prior to account for LAC global scaling. Furthermore, the bias versus background variability (surrogate for noise) tradeoff for sTX-MLAA PET images was greatly improved compared to correcting for attenuation using coincidences segmented from the TX source alone (Fig. 5). Thus, use of patient coincidences in sTX-MLAA mitigated sparse sampling artifacts, when reconstructing μ-maps with external source events only. In sum, sTX-MLAA is a compromise approach that produces PET images with a higher accuracy than MLAA, enabled by using coincidences from the TX source, and lower noise than using TX source generated μ-maps, afforded by leveraging the typically increased count rate due to tracer in the patient.

Uptake bias in the activity-free lung insert, for experiments with the NEMA IEC body phantom, was ≥7% for sTX-MLAA versus CT-based AC PET images (Fig. 3d and Fig. 6c). A source of this error was limited tomographic sampling for the sparse TX source. The initial μ-map for sTX-MLAA (Algorithm 1) was produced by transmission reconstruction of segmented external source counts alone (line 4 in Algorithm 2), which used a starting μ-map of LACs of uniform water. Thus, voxels with low tomographic sampling in the initial TX reconstruction would overestimate LACs, which was unresolved with sTX-MLAA iterations. We note that residual activity is measured in the lung insert, even for PET images reconstructed with CT-AC [22]. Thus, lung LAC overestimation amplifies existing residual uptake. A second cause was cross-talk between emission and transmission updates, due to the residual lung uptake itself. Lung residual has been found to decrease in CT-AC results as a function of improved TOF resolution; compare the Siemens Biograph mCT [22] (TOF resolution=580 ps, lung residual=12.1%) with the Vision [23] (TOF resolution=210 ps, lung residual=3.5%), such that cross-talk in sTX-MLAA would also decrease. Although we observed uptake bias >5% in the lung insert for sTX-MLAA, bias was only 0.7% for the lung compartment filled with an FDG activity concentration comparable to patient imaging in the phantom with a cardiac insert (Fig. 8c). Thus, for human FDG imaging we expect high lung uptake accuracy.

Despite promising results, there are several areas for improvement of sTX-MLAA. A key challenge is addressing noise amplification in PET images compared to CT-AC. Increased noise for sTX-MLAA is due to 1) higher randoms from addition of the TX source (not evaluated here), and 2) noise propagation to PET images from the reconstructed μ-map (increases with $\alpha$). External source activity at the start of imaging was 12.7-16.5 MBq. Our rationale for this strength was empirically determined by preliminary experiments and prior reports [9], and was expected to prioritize PET bias over noise amplification. Future studies will explore optimizing source strength. We used a penalized-likelihood transmission algorithm to improve the bias-noise tradeoff for μ-maps, but further noise reduction is needed. For two phantoms (NEMA without arms, and cardiac insert) optimal $\alpha$ was ~10 while it was ~1 for the IEC phantom with arms. The TX source coincidence count rate increases towards the edge of sinogram projection bins, such that a lower $\alpha$ may have been sufficient to correct for MLAA limitations for the IEC phantom with arms. Ideally, this parameter should be adjusted based on the PET data alone. Finally, we used absolute scatter estimation for correcting counts from the patient. The method underestimates scatter from out of FOV [16], which may bias µ-map estimation at edge planes; increasing PET bias of extracardiac disease.

TABLE III
MEASURED MEAN UPTAKE ACROSS ORGANS FOR THE ANTHROPOMORPHIC TORSO PHANTOM WITH CARDIAC INSERT

| Organ | Uptake (kBq/ml) | | Diff. (%)[a] |
|---|---|---|---|
| | CT | sTX-MLAA | |
| Background | 1.8 | 1.8 | -0.2 |
| Liver | 4.8 | 5.0 | 3.7 |
| Lungs | 0.8 | 0.8 | -0.7 |
| Cardiac Defect | 5.4 | 5.5 | 0.9 |
| Ventricular Blood | 3.7 | 3.7 | -0.7 |

[a]normalized to CT-AC PET images

The key benefit of sTX-MLAA is the ability to robustly estimate and correct for attenuation using data acquired simultaneously during the PET exam where AC will be applied, without use of prior results or a µ-map intensity prior (i.e. to account for LAC global scaling). This could have a substantial benefit on PET artifacts due to patient motion mismatch between the µ-map and PET acquisitions. We note that cardiac MR exams at our institution average 60 min (interquartile range: 47-79 min). Thus, if PET data were collected throughout a PET/MR study, we would expect substantial PET respiratory and non-cyclical body motion artifacts [24] when applying MR-AC generated with MR images acquired at a single timepoint. By using sTX-MLAA for AC, patient motion in the µ-map will exactly match that in tracer images; eliminating motion mismatch errors.

## V. CONCLUSION

This paper presents a new AC strategy that uses a physically fixed and sparse transmission source that is relatively easy to fill and place. The method combines the SNR benefits of conventional MLAA, using coincidences from the patient, with the higher accuracy of transmission reconstruction with an external source. AC with sTX-MLAA produced PET images with ROI uptake within 5% of CT-AC results for phantom scans. Noise and sparse sampling artifacts were largely reduced versus AC using segmented coincidences from the external source alone. Results suggest that sTX-MLAA will enable quantitative PET during cardiac PET/MR of human patients.

## ACKNOWLEDGEMENT

The authors thank Michael Casey, Vladimir Panin, Harold Rothfuss, Deepak Bharkhada, and David LaFratta, from Siemens Medical Solutions, for useful discussions and technical guidance. The authors also acknowledge Orhan Öz, Jon Anderson, and Nima Kasraie, from UT Southwestern Medical Center, for useful discussions. Funding was provided in part from the National Institutes of Health (grant number NIH-NIBIB R03EB028946-01A1).

## REFERENCES

[1] T. Vita et al., "Complementary Value of Cardiac Magnetic Resonance Imaging and Positron Emission Tomography/Computed Tomography in the Assessment of Cardiac Sarcoidosis," Circ. Cardiovasc. Imaging, vol. 11, no. 1, p. e007030, Jan. 2018.
[2] A. Lebasnier et al., "Diagnostic value of quantitative assessment of cardiac F-18-fluoro-2-deoxyglucose uptake in suspected cardiac sarcoidosis," Ann. Nucl. Med., vol. 32, no. 5, pp. 319–327, Jun. 2018.
[3] A. Ahmadian et al., "Quantitative interpretation of FDG PET/CT with myocardial perfusion imaging increases diagnostic information in the evaluation of cardiac sarcoidosis," J. Nucl. Cardiol., vol. 21, no. 5, pp. 925–939, Oct. 2014.
[4] V. Vergani et al., "Hybrid cardiac PET/MR: the value of multiparametric assessment in cardiac sarcoidosis," Clin. Transl. Imaging, vol. 7, no. 5, pp. 317–326, Oct. 2019.
[5] P. M. Robson et al., "MR/PET Imaging of the Cardiovascular System," JACC-Cardiovasc. Imaging, vol. 10, no. 10, pp. 1165–1179, Oct. 2017.
[6] M. L. Lassen et al., "Assessment of attenuation correction for myocardial PET imaging using combined PET/MRI," J. Nucl. Cardiol., vol. 26, no. 4, pp. 1107–1118, 2019.
[7] A. Rezaei, C. M. Deroose, T. Vahle, F. Boada, and J. Nuyts, "Joint Reconstruction of Activity and Attenuation in Time-of-Flight PET: A Quantitative Analysis," J. Nucl. Med., vol. 59, no. 10, pp. 1630–1635, Oct. 2018.
[8] M. Teimoorisichani, V. Panin, H. Rothfuss, H. Sari, A. Rominger, and M. Conti, "A CT-less approach to quantitative PET imaging using the LSO intrinsic radiation for long-axial FOV PET scanners," Med. Phys., vol. 49, no. 1, pp. 309–323, 2022.
[9] P. Mollet, V. Keereman, J. Bini, D. Izquierdo-Garcia, Z. A. Fayad, and S. Vandenberghe, "Improvement of attenuation correction in time-of-flight PET/MR imaging with a positron-emitting source," J. Nucl. Med., vol. 55, no. 2, Art. no. 2, Feb. 2014.
[10] V. Y. Panin, M. Aykac, and M. E. Casey, "Simultaneous reconstruction of emission activity and attenuation coefficient distribution from TOF data, acquired with external transmission source," Phys. Med. Biol., vol. 58, no. 11, Art. no. 11, Jun. 2013.
[11] S. L. Bowen, N. Fuin, M. A. Levine, and C. Catana, "Transmission imaging for integrated PET-MR systems," Phys. Med. Biol., vol. 61, no. 15, Art. no. 15, Aug. 2016.
[12] S. L. Bowen, "Optimization of Supplemental Transmission Source Imaging for Joint Transmission-Emission Scanning on an Integrated PET-MR System," in IEEE Nucl. Sci. Symp. Conf., 2016.
[13] H. Erdogan and J. A. Fessler, "Ordered subsets algorithms for transmission tomography," Phys. Med. Biol., vol. 44, no. 11, Art. no. 11, Nov. 1999.
[14] L. Cheng, T. Ma, X. Zhang, Q. Peng, Y. Liu, and J. Qi, "Maximum likelihood activity and attenuation estimation using both emission and transmission data with application to utilization of Lu-176 background radiation in TOF PET," Med. Phys., vol. 47, no. 3, pp. 1067–1082, Mar. 2020.
[15] C. C. Watson, M. E. Casey, C. Michel, and B. Bendriem, "Advances in scatter correction for 3D PET/CT," in 2004 IEEE Nuclear Science Symposium Conference Record, Vols 1-7, J. A. Seibert, Ed. New York: IEEE, 2004, pp. 3008–3012.
[16] H. Bal, J. W. Kiser, M. Conti, and S. L. Bowen, "Comparison of maximum likelihood and conventional PET scatter scaling methods for 18 F-FDG and 68 Ga-DOTATATE PET/CT," Med. Phys., vol. 48, no. 8, pp. 4218–4228, Aug. 2021.
[17] L. G. Byars et al., "Variance reduction on randoms from coincidence histograms for the HRRT," in IEEE Nuclear Science Symposium Conference Record, 2005, Oct. 2005, vol. 5, pp. 2622–2626.
[18] M. E. Casey, H. Gadagkar, and D. Newport, "Component Based Method for Normalization in Volume PET," Int. Meet. Fully Three-Dimens. Image Reconstr. Radiol. Nucl. Med., 1995.
[19] T. Norikane, Y. Yamamoto, Y. Maeda, T. Noma, H. Dobashi, and Y. Nishiyama, "Comparative evaluation of F-18-FLT and F-18-FDG for detecting cardiac and extra-cardiac thoracic involvement in patients with newly diagnosed sarcoidosis," EJNMMI Res., vol. 7, p. 69, Aug. 2017.
[20] National Electrical Manufactures Association, NEMA Standards Publication NU-2-2007: Performance Measurements of Positron Emission Tomographs. Rosslyn, VA: National Electrical Manufactures Association, 2007.
[21] R. A. Coulden, E. P. Sonnex, J. T. Abele, and A. M. Crean, "Utility of FDG PET and Cardiac MRI in Diagnosis and Monitoring of Immunosuppressive Treatment in Cardiac Sarcoidosis," Radiol. Cardiothorac. Imaging, vol. 2, no. 4, p. e190140, Aug. 2020.
[22] B. W. Jakoby, Y. Bercier, M. Conti, M. E. Casey, B. Bendriem, and D. W. Townsend, "Physical and clinical performance of the mCT time-of-flight PET/CT scanner," Phys. Med. Biol., vol. 56, no. 8, Art. no. 8, Apr. 2011.
[23] J. van Sluis et al., "Performance Characteristics of the Digital Biograph Vision PET/CT System," J. Nucl. Med., vol. 60, no. 7, pp. 1031–1036, Jul. 2019.
[24] T. Sun et al., "Body motion detection and correction in cardiac PET: Phantom and human studies," Med. Phys., vol. 46, no. 11, pp. 4898–4906, 2019.